\documentclass[12pt,a4paper,reqno]{amsart2000}

\def\a{\alpha}
\def\b{\beta}

\def\e{\epsilon}

\def\l{\lambda}
\def\o{\omega}

\def\F{\Phi}
\def\G{\Gamma}

\def\O{\Omega}

\font\bss=cmssbx10 at 12 pt

\def\bC{{\mathbb C}}
\def\bR{{\mathbb R}}

\def\bN{{\mathbb N}}

\newfont{\goth}{eufm10 scaled \magstep1}

\def\gg{\mbox{\goth g}}

\def\gs{\mbox{\goth s}}

\def\gX{\mbox{\goth X}}
\def\gS{\mbox{\goth S}}

\def\Sp#1{{\mathrm{Sp(#1)}}}

\newfont{\mcal}{eusm10 scaled \magstep1}

\def\cd{\mbox{\mcal D}}

\def\cu{\mbox{\mcal U}}

\def\fr#1#2{{\textstyle{\frac{#1}{#2}}}}
\def\p{\partial}

\def\square{\kern1pt\vbox
                {\hrule height 0.6pt\hbox{\vrule width 0.6pt\hskip 3pt
     \vbox{\vskip 6pt}\hskip 3pt\vrule width 0.6pt}\hrule height 0.6pt}\kern1pt}

\def\La{\wedge}
\def\ra{\rightarrow}

\def\End{\mathrm{End\;}}

\def\wt{\widetilde}

\newtheorem{Th}{Theorem}
\newtheorem{Prop}{Proposition}
\newtheorem{Cor}{Corollary}
\newtheorem{Lem}{Lemma}
\newtheorem{Def}{Definition}

\def\bt{\begin{Th}\ \ }
\def\et{\end{Th}}
\def\bp{\begin{Prop}\ \ }
\def\ep{\end{Prop}}
\def\bc{\begin{Cor}\ \ }
\def\ec{\end{Cor}}
\def\bl{\begin{Lem}\ \ }
\def\el{\end{Lem}}
\def\bd{\begin{Def}\ \ }
\def\ed{\end{Def}}

\def\qed{\hfill\square}
\def\n{\nabla}
\def\op{\oplus} \def\ot{\otimes}
\def\dpp{\p_{++} } \def\App{A_{++} } 
\def\dmm{\p_{--} }  
\def\dpm{\p_{\pm\pm} }

\def\be{\begin{equation}}
\def\ee{\end{equation}}
\def\la#1{\label{#1}}
\def\re#1{(\ref{#1})}
\def\arr{\begin{array}{rlll}}
\def\ea{\end{array}}
\def\bea{\begin{eqnarray}}
\def\eea{\end{eqnarray}}
\def\bean{\begin{eqnarray*}}
\def\eean{\end{eqnarray*}}

\def\3s{3-Sasakian manifold}
\def\hk{hyper-K\"ahler}

\def\sdy{self-duality}
\def\sd{self-dual}
\def\ym{Yang-Mills}
\def\hf{half-flat} \def\gs{Grassmann structure}

\begin{document}
\rightline{hep-th/0205030}
\vskip 1.5 true cm

\title[Partially-flat gauge fields]{Partially-flat gauge fields on manifolds 
of dimension greater than four}
 
\author[D.V. Alekseevsky, V. Cort\'es]
{Dmitri V.\ Alekseevsky, Vicente Cort\'es} 
\address{Dept. of Mathematics\\
University of Hull\\
Hull, HU6 7RX\\
UK}
\email{D.V.Alekseevsky@maths.hull.ac.uk}
\author[C. Devchand]{Chandrashekar Devchand}
\address{Mathematisches Institut\\
Universit\"at Bonn\\
Beringstr. 1\\ 
D-53115 Bonn}
\email{vicente@math.uni-bonn.de} 
\email{devchand@math.uni-bonn.de}
\thanks{
Talk given by C.D.\ on 11th September, 2001
at the Workshop on Special Geometric Structures in String Theory, Bonn.
This work was supported by the `Schwerpunktprogramm Stringtheorie' of the
Deutsche For\-schungs\-ge\-mein\-schaft.} 
\begin{abstract}
We describe two extensions of the notion of a self-dual connection 
in a vector bundle over a manifold $M$ from $\dim M{=}4$ to higher dimensions. 
The first extension, $\O$-self-duality, is based on the
existence of an appropriate 4-form $\O$ on the Riemannian manifold $M$ and
yields solutions of the Yang-Mills equations. The second is the
notion of half-flatness, which is defined for manifolds with certain
Grassmann structure $T^{\bC}M {\cong}  E \otimes H$. In some cases,
for example for hyper-K\"ahler manifolds $M$, half-flatness implies
$\O$-self-duality. A construction of half-flat connections inspired by 
the harmonic space approach is described.
Locally, any such connection can be obtained from a free prepotential
by solving a system of linear first order ODEs. 
\end{abstract}
\maketitle

\section{Self-duality and half-flatness}

The Yang-Mills self-duality equations have played a very important role
in field theory and in differential geometry. This talk is about ongoing
work on two generalisations of the notion of {\it self-duality} from four to
higher dimensional manifolds $M$. The first generalisation is the notion of

\noindent
{\bss ${\mathbf \O}$-self-duality}.

\noindent
This is based on the existence of an appropriate 4-form $\O$ on a
pseudo-Riemannian manifold $(M,g)$.
For $\O \in \O^4(M)$ we define a traceless symmetric
endomorphism field $B_\O : \La^2 T^*M \ra \La^2 T^*M$ by
\be
B_\O \o := \ast  (\ast  \O \wedge \o )\ ,
\ee
where $\o\in \La^2 T^*M$.
\bd
A 4-form $\ \O \in \O^4(M)\,$ on a pseudo-Riemannian manifold $M$ is
called {\bss appropriate} if there exists a non-zero real constant
eigenvalue $\l$ of the endomorphism field  $B_\O$.
\la{approp}
\ed
\noindent
We note that on a Riemannian manifold the eigenvalues of $B_\O$ are
real for any 4-form $\O$.
We can now define a generalisation of the four dimensional notion
of self-duality thus:
\bd
Let $\O$ be an appropriate 4-form on a pseudo-Riemannian manifold $(M,g)$
and $\l\neq 0 \in \bR$. A connection $\n$ in a vector bundle $\nu: W\ra M$
is {\bss ${\mathbf  \O}$-self-dual} if its curvature $F^\n$ satisfies
the linear algebraic system
\bea 
B_\O F^\n &=& \l F^\n  \la{sdy1}\\[2pt]
(d\ast \O) \wedge F^\n &=& 0\ .
\la{sdy2}
\eea
\ed
\noindent
It is easy to see that an $\O$-self-dual connection $\n$ is a \ym\
connection, i.e. it satisfies the \ym\ equation:
$$ d^\n \ast F^\n= \pm \fr{1}{\l}  d^\n\left(\ast \O\wedge F^\n\right) 
= \pm \fr{1}{\l} \left((d\ast \O) \wedge F^\n 
                    + \ast \O\wedge d^\n F^\n\right)= 0 
$$
in virtue of eq.\ \re{sdy2} and the Bianchi identity $ d^\n  F^\n = 0$.
This extension of \sdy\ is based on early work of \cite{CDFN}
on gauge fields in flat spaces. Interesting examples of manifolds
with appropriate (and parallel) 4-forms are provided, for instance by 
manifolds with special holonomy groups.
So for gauge fields on these manifolds, $\O$-self-duality may be
defined. Various examples have been discussed in the literature, e.g. in
\cite{CS,BKS,DT,T}.

The second generalisation of \sdy\ is the notion of

\noindent
{\bss Half-flatness}.

\noindent
This is defined for manifolds with
certain Grassmann structure. In what follows $M$
denotes a complex manifold. A \gs\ on $M$ is an isomorphism of the
(holomorphic) tangent bundle, $TM \cong E\ot H$, where $E$ and $H$ are two
holomorphic vector bundles
over $M$. This is just a generalised spinor decomposition, allowing
representation of vector fields using two types of spinor indices, viz.
$X_{a\a}:= e_a \ot h_\a$, where $(e_a)$ and $(h_\a)$ are frames of
$E$ and $H$ respectively. If $M$ is four dimensional, both $E$ and $H$
have rank 2 and $a,\a$ are the familiar 2-spinor indices. Manifolds with
\gs\ provide interesting generalisations of four dimensional manifolds.
A  Grassmann connection $\n$ is a linear connection which preserves
the \gs:
$$
\n (e\ot h) = \n^E e \ot h + e \ot \n^H h\, ,
$$
where $\n^E,\n^H$ are connections in the bundles $E,H$ and $e,h$ are
local sections of $E,H$ respectively. A Grassmann structure with
Grassmann connection $\n$ is called
half-flat if the connection $\n^H$ in $H$ is flat.
A manifold with such a half-flat  Grassmann structure
is called a half-flat Grassmann manifold. Here 
we assume that rank $H=2$ and that a $\n^H$-parallel non-degenerate
fibre-wise
2-form $\o_H \in \G(\La^2 H^*)$ in the bundle $H$ is fixed, so these
manifolds are generalisations of \hk\ manifolds. The torsion of
the Grassmann connection lives in the space
\bean
TM\ot \La^2 T^*M &=&
TM \ot \left(S^2H^* \ot \La^2 E^*\ \op\ \La^2 H^*  \ot S^2E^* \right)
\\[6pt]
           &=& EH \left( S^2H^* \La^2 E^*\ \op\  \o_H S^2 E^* \right)
\\[6pt] &\cong& \left( S^3 H \ \op\  \o_H H \right)E\La^2 E^*
                                \ \op\  \o_H H E S^2 E^*\, ,
\eean
where we omit the $\ot$'s and we identify $H^*$ with $H$ using
$\o_H$. Notice that since the bundle $H$ has rank 2, the line bundle
$\La^2 H$ is generated by the 2-form $\o_H$.
(Choosing a frame $(h_1,h_2)$ for $H$, we may write
$\o_H(h_\a,h_\b) := \e_{\a\b}$, a skew matrix with $\e_{12}=1$.)

We now consider gauge fields on these manifolds.
Let $\n$ be a connection in a holomorphic vector bundle $\nu: W\ra M$.
(If $\nu$ has structure group $G$,
we may choose a frame for it, in which the vector potential takes values
in the gauge Lie algebra $\gg$). The curvature of the connection $\n$ is
a 2-form with values in the gauge algebra,
$ F^\n \in \La^2 T^*M \ot \End W $. Now, as above,
in virtue of the \gs\ the
space of bivectors decomposes as
$$ \La^2 TM = S^2 E \ot \La^2 H  \oplus \La^2 E  \ot S^2 H\ =
S^2 E \ot \o_H  \oplus \La^2 E  \ot S^2 H\, .
$$
If  $e,e'\in \G(E)$, we have the decomposition
$$ F(e\ot h_\a\,,\, e'\ot h_\b) = \o_H(h_\a,h_\b) F^{(e,e')} +
F^{[e,e']}_{(\a \b )}\ ,
$$
where $F^{(e,e')}$ is  symmetric in $e,e'$
and $F^{[e,e']}_{(\a \b )}$ is skew in $e,e'$
and symmetric in $\a \b$.
In 4 dimensions both $E$ and $H$ have rank 2, so this corresponds
to the familiar decomposition into self-dual and anti-\sd\ parts
of the curvature.

\bd
Let $M$ be a manifold with \gs , not necessarily half-flat.
A connection $\n$ in a vector
bundle $W\ra M$  is called {\bss half-flat} if its curvature
$$
F^\n \in \La^2 H \ot S^2 E \ot \End W   = \o_H \ot S^2 E \ot \End W\ .
$$
\ed
Quaternionic K\"ahler and \hk\ manifolds are special cases of manifolds
with (a locally defined) \gs , the Levi-Civita connection being the
canonical Grassmann connection. In the \hk\ case the
Grassmann structure is half-flat. For these examples, half-flatness of a
gauge connection implies $\O$-\sdy\, with respect to the canonical 
parallel 4-form $\O$. In other words, a \hf\ curvature is one of
the eigenstates of the endomorphism field $B_\O$, as in 4 dimensions.
Therefore, in these cases, the two generalisations
of the idea of \sdy\ mentioned above coincide.

There is an interesting special case of \gs, considered
by \cite{AG}:
\bd
A {\,\bss spin $\boldsymbol{\fr{m}{2}}\,$ Grassmann structure\,} on a
(complex) manifold $M$ is a holomorphic Grassmann structure of the form
$TM \cong E\ot F =  E\ot S^m H$, with a holomorphic Grassmann connection
$\n^{TM} = \n^E \ot {\rm Id} + {\rm Id} \ot \n^F$, where $H$ is a rank 2
holomorphic vector bundle over $M$ with symplectic form $\o_H$ and
holomorphic symplectic connection $\n^H$, and $\n^F$ is the connection
in $F=S^mH$ induced by $\n^H$.
$M$ is called {\bss half-flat spin $\boldsymbol{\fr{m}{2}}\,$ Grassmann
manifold} if the connection $\n^F$ is flat.
\ed
The bundle $S^m H$ is associated with the
spin $\fr{m}{2}$ representation of the group $\Sp{1,\bC}$.
Any frame $(h_1,h_2)$ for $H$ defines a frame for $S^m H$,
$(h_A := h_{\a_1}h_{\a_2}\dotsm h_{\a_m})$, where the multi-index
$A:=\a_1\a_2\dots\a_m\ ,\ \a_i=1,2$.
The symplectic form $\o_H$ induces a bilinear form $\o_H^m$ on $F=S^m H$
given by
$$
\o_H^m(h_A,h_B) := \gS_A \gS_B \o_H(h_{\a_1},h_{\b_1})
        \o_H(h_{\a_2},h_{\b_2})\dotsm \o_H(h_{\a_m},h_{\b_m})\ ,
$$
where $\gS_A$ denotes the sum over all permutations of the $\a$'s.
This form is skew-symmetric if $m$ is odd and symmetric if $m$ is even.
To any section $e\in \G(E)$ and multi-index $A$ we associate
the vector field $X^e_A := e\ot h_A$ on $M$ and we put $X_{aA} := X_A^{e_a}$.

Let $(M,\n^{\rm TM})$ be a half-flat spin $\fr{m}{2}$ Grassmann manifold
and $\n$ a connection in a holomorphic vector bundle $W\ra M$.
The  space of bivectors $\La^2 TM$ has the following decomposition
into $GL(E)\ot \Sp{1,\bC}$-submodules:
$$ \La^2 TM = \La^2\Bigl( E\ot S^mH\Bigr)
                  = \La^2E \ot S^2 S^mH \ \op\  S^2E  \ot  \La^2S^mH\, ,
$$
where
\bean S^2 S^mH &=& S^{2m}H\ \op\ \o_H^2 S^{2m-4}H\ \op\ \dotsb  \op
\o_H^{2[\frac{m}{2}]} S^{2m -4[\frac{m}{2}]}H\\[8pt]
\La^2S^mH &=&  \o_H S^{2m-2}H\ \op\  \o_H^3 S^{2m-6}H \op \dotsb
\op \o_H^{2[\frac{m}{2}]+1} S^{2m - 4[\frac{m}{2}]-2}H\, .
\eean
Here we use the convention that $S^lH = 0$ if $l< 0$.
The corresponding decomposition of the curvature of $\n$ is:
\bean
&& F\bigl(X^e_A, X^{e'}_B\bigr) \nonumber\\
&& =\ \gS_A \gS_B \sum_{k=0}^{\left[m/2\right]}
         \bigl(
         \o_H(h_{\a_1},h_{\b_1})\dotsm \o_H(h_{\a_{2k}},h_{\b_{2k}})
    \stackrel{(2k)}{F}{}^{[ee']}_{\a_{2k+1}\dots\a_m \b_{2k+1}\dots\b_m }
\bigr.\nonumber\\
&&\qquad\qquad + \bigl.
         \o_H(h_{\a_1},h_{\b_1})\dotsm \o_H(h_{\a_{2k+1}},h_{\b_{2k+1}})
  \stackrel{(2k+1)}{F}{}^{(ee')}_{\a_{2k+2}\dots\a_m \b_{2k+2}\dots\b_m}
\bigr)\ ,
\la{curv_decomp}\eean
where the tensors $\stackrel{(2k)}{F}\in \G(\La^2 E^* \ot S^{2m-4k}H^*)$
and  $\stackrel{(2k+1)}{F}\in \G(S^2 E^* \ot S^{2m-4k-2}H^*)$
are the curvature components in irreducible
$GL(E){\cdot}\Sp{1,\bC}$-submodules.

For manifolds with spin $\fr{m}{2}$ Grassmann structure it is natural
to define half-flat connections as those satisfying
$$
\stackrel{(2i)}{F} = 0\ ,\quad {\rm for\ all}\ i\in \bN .
$$
However, for these manifolds it is useful to consider a more
refined notion:

\bd
A connection $\n$ in the vector bundle $\nu: W\ra M$ is called
{\bss k-partially flat} if $\stackrel{(i)}{F}=0$ for all $i\le 2k$.
Here $0\le k\le [\fr{m+2}{2}]$.
\ed
\noindent
Clearly, $[\fr{m+2}{2}]$-partially flat connections are simply flat
connections. We note that for $m{=}1\,$, 0-partially flat connections
are precisely half-flat connections.
For general odd $m{=}2p{+}1\,$, 0-partially flat connections in a
vector bundle $\nu$ over flat spaces with spin $\fr{m}{2}$
Grassmann structure were considered in \cite{W}. Some other k-partially
flat connections on flat spaces were considered in \cite{DN}.

The penultimate case $k = [\fr{m}{2}]$ is particularly interesting for
odd $m$.
\bt Let $M$ be a  half-flat
spin $\fr{m}{2}$ Grassmann manifold $M$. If $m$ is odd and the
vector bundle $E\ra M$ admits a $\,\n^E$--parallel symplectic form
$\o_E$, then $M$ has canonical $\,\Sp{E}{\cdot}\Sp{H}$--invariant metric
$\,g = \o_E \otimes \o_H^m\,$ and $4$-form $\Omega \neq 0$.
If $\,\O\,$ is co-closed with respect to the metric $g$, then any
$\fr{m-1}{2}$-partially flat connection $\n$ in a vector bundle
$W$ over $M$ is $\O$-selfdual and hence it is a Yang-Mills connection,
i.e. it satisfies the \ym\ equation.
\et

\noindent
{\it Sketch of proof:}
To describe $\O$ we use the following notation: $e_a$ is a basis of $E$,
$h_\a$ is a basis of $H$, $h_A$ is the corresponding basis of $S^mH$ and
$X_{aA} := e_a\ot h_A$ the corresponding basis of $TM = E\ot S^mH$.
With respect to these bases, the skew symmetric forms $\o_E$, $\o_H$ and
$\o_H^m$ are represented by the matrices $\o_{ab}$, $\o_{\a \b }$ and
$\o_{AB}$ respectively. We define $\O$ by
\[ \O := \sum \o_{ab}\o_{cd}\o_{AC}\o_{BD}X^{aA}\wedge X^{bB}\wedge
X^{cC}\wedge X^{dD} \, ,\]
where $X^{aA}$ is the dual basis of the basis $X_{aA}$.
The connection $\n$ is $\fr{m-1}{2}$-partially flat if and only if its
curvature $F$ belongs to the space
$ S^2E^* \otimes \o_H^m \otimes \End (W)$. Contracting
a tensor $S = S_{ab}\o_{AB}e^ae^b\otimes h^Ah^B$ in $S^2E^* \otimes \o_H^m$
with $\O$, after some algebra, yields $\l S $ with $\l = -4(m+1) \neq 0$.
Hence any $\fr{m-1}{2}$-partially flat connection
is $\O$-selfdual and is a Yang-Mills connection.
\qed

\section{Construction}
We now briefly sketch a construction of \hf\ connections on \hf\ Grassmann
manifolds using a variant of the {\it harmonic space} approach 
(c.f.\ \cite{GIOS1}).
The method affords application to a construction of partially flat
connections on a  half-flat
spin $\fr{m}{2}$ Grassmann manifold as well.
Details and proofs will appear in \cite{ACD}.

We denote by $S_H$ the $\Sp{1,\bC}$-principal holomorphic
bundle over $M$
consisting of symplectic bases of $H_m \cong \bC^2\,,\ m\in M$,
$$
S_H=\{s=(h_+,h_-)\;|\;\o_H(h_+,h_-)=1 \}\, .
$$
The bundle $S_H\rightarrow M$ is called {\bss harmonic space}.
A fixed parallel symplectic frame $(h_1,h_2)$ of $H$, such that
$ m \mapsto s_m=(h_1(m),h_2(m))\in S_H$ yields a trivialisations
$M \times\Sp{1,\bC} \cong S_H\ $ defined by
$$
(m, \cu ) \mapsto s_m \cu
= \left(h_+ = \sum h_\a u_+^\a\;,\; h_- = \sum h_\a u_-^\a\right),\,
\cu= \begin{pmatrix} u_+^1 & u_-^1 \\  u_+^2 & u_-^2 \end{pmatrix};\,
                   \det \cu = 1 \, .
$$
This is precisely the harmonic space of \cite{GIKOS, GIOS2}. The matrix
coefficients $u^\a_\pm$ of  $\Sp{1,\bC}$ are considered as holomorphic
functions on $S_H$ and together with local coordinates of $M$, we
obtain a system of local coordinates on $S_H$ constrained by the
relation $u^1_+u^2_- -u^1_-u^2_+ = 1$.
We denote by  $\dpp,\dmm,\p_0$ the fundamental vector fields on $S_H$
generated by the standard generators of $\Sp{1,\bC}$,
satisfying the relations
$$
\left[ \dpp\,,\,\dmm \right] = \p_0\ ,\quad
\left[ \p_0\,,\,\dpp \right] = 2 \dpp\ ,\quad
\left[ \p_0\,,\,\dmm \right] = -2\dmm\ .
$$
For any section $e\in \G(E)$ we define vector fields $X_\pm^e \in \gX(S_H)$
by the formula
$$ X_\pm^e \vert_{(h_+,h_-)} = \wt{e\ot h_\pm}\ ,$$
where $\wt{Y}$ is the horizontal lift of a vector field $Y$ on $M$.
They satisfy the relations
\bean
&&\left[\p_0, X^e_\pm \right] = \pm X^e_\pm\ ,\quad
\left[\dpm, X^e_\pm \right] = 0\ ,\quad
\left[\dpm, X^e_\mp \right] = X^e_\pm\ ,
\\[5pt]
&&\bigl[X_\pm^e,X_\pm^{e'}\bigr]
\ =\  X_\pm^{\n_{\pi_*X^e_\pm} e'} - X_\pm^{\n_{\pi_*X^{e'}_\pm} e }
          - \wt{T}( \pi_* X^e_\pm, \pi_* X^{e'}_\pm)\,,
\\[5pt]
&&\bigl[X_+^e,X_-^{e'}\bigr]
\ =\   X_-^{\n_{\pi_*X^e_+} e'}  - X_+^{\n_{\pi_*X^{e'}_-} e }
          - \wt{T}( \pi_* X^e_+, \pi_* X^{e'}_-)\, ,
\eean
where $T$ is the torsion of the Grassmann connection, 
$\wt{T}(X,Y) :=\wt{T(X,Y)}$ denotes the horizontal lift of the vector  
$T(X,Y)$ and we have used the abbreviation $\n_Xe := \n_X^Ee\,$. 
We denote by $\cd_\pm$ the distributions generated by vector fields of 
the form $X^e_\pm\,$, $e\in  \G(E) $.

The holomorphic \gs\ is called  admissible
if the torsion of the Grassmann connection $\n$ has no component in
$S^3 H \ot E\ot \La^2 E^*$. This means, in particular, that the torsion
can be written as a sum of tensors linear in $\o_H$.
Now, it is not difficult to prove  that if a \gs\ is  admissible,
the distribution $\cd_+$ (associated to any parallel frame $(h_1, h_2)$)
on $S_H$ is integrable.

The basic idea of the harmonic space construction is to lift the
geometric data from $M$ to $S_H$ via the projection $\pi: S_H\ra M$.
The pull back  $\pi^* \n$ of a half-flat connection $\n$ in the trivial 
vector bundle $\nu: W = \bC^r \times M \ra M$ is a connection in the vector 
bundle $ \pi^*\nu: \pi^* W\ra S_H$ which satisfies equations defining the 
notion of a half-flat gauge connection on $S_H$. One can also define the weaker
notion of an almost half-flat connection in $ \pi^*\nu: \pi^* W\ra S_H$
by considering only the equations on the curvature involving  
$\p_0$, $\dpp$ and $X_+^e$ in one of the arguments.

Now, the main result, for
which the integrability of $\cd_+$ is crucial, can be stated:

\bt
Let $M$ be a manifold with a half-flat admissible Grassmann
structure. Consider a matrix-valued function $A_{++}\ $ on $S_H$ 
(the ``analytic prepotential''), constant along the leaves
of $\cd_+$ with the homogeneity property $\p_0 A_{++}=2 A_{++}\ $, 
and an invertible matrix-valued solution of the system of first-order
linear ODE's, $\dpp \F = - \App \F\ ,\  \p_0 \F =0\,$. The pair $(\App,\F)$
defines an almost half-flat connection $\n^{(\App,\F)}$ in the bundle
$ \pi^*\nu: \pi^* W\ra S_H$, which depends only on $\App,\F$ and their
first and second partial derivatives. The potential $A(X_+^e)$ of this 
connection with respect to the frame  $e\F= e_i \F^i_j$, where
$e=(e_i)$ is the standard frame of $W = \bC^r \times M$, 
determines a  half-flat connection in the bundle $ \nu: W\ra M$.
Conversely, any half-flat connection over $M$ can be obtained using
this construction.
\et
\noindent
The proof and further details may be found in \cite{ACD}.

\end{document}